%% file: main.tex
\documentclass[11pt]{article}

\usepackage{cayley}

\title{Sparsifying Cayley Graphs on Every Group}
\date{\today}

\author{
    Jun-Ting Hsieh\thanks{MIT. \texttt{juntingh@mit.edu}.}
    \and Daniel Z. Lee\thanks{MIT.  \texttt{lee\_d@mit.edu}. Supported by the NSF CAREER grant CCF-2443045, and the
Reed Fund at MIT.}
    \and Sidhanth Mohanty\thanks{MIT. \texttt{sidm@mit.edu}.
    Supported by NSF Award DMS-2022448.}
    \and Aaron Putterman\thanks{Harvard University. \texttt{aputterman@g.harvard.edu}. Supported in part by the Simons Investigator Awards of Madhu Sudan and Salil Vadhan, NSF Award CCF 2152413 and AFOSR award FA9550-25-1-0112.}
    \and Rachel Yun Zhang\thanks{MIT. \texttt{rachelyz@mit.edu}. Supported by NSF Graduate Research Fellowship 2141064. Supported in part by NSF grant CNS-2154149.}
}

\begin{document}

\maketitle

\begin{abstract}
\input{content/abstract}
\end{abstract}

\thispagestyle{empty}
\setcounter{page}{0}
\newpage


\input{content/intro}

\input{content/tech-overview}

\input{content/sparsification}

\bibliography{ref}
\bibliographystyle{alpha}

\end{document}

%% file: content/abstract.tex
A classic result in graph theory, due to Batson, Spielman, and Srivastava (STOC 2009)  shows that every graph admits a $(1 \pm \varepsilon)$ cut (or spectral) sparsifier which preserves only $O(n / \varepsilon^2)$ reweighted edges. However, when applying this result to \emph{Cayley graphs}, the resulting sparsifier is no longer necessarily a Cayley graph --- it can be an arbitrary subset of edges.

Thus, a recent line of inquiry, and one which has only seen minor progress, asks: for any group $G$, do all Cayley graphs over the group $G$ admit sparsifiers which preserve only $\mathrm{polylog}(|G|)/\varepsilon^2$ many re-weighted generators?

As our primary contribution, we answer this question in the affirmative, presenting a proof of the existence of such Cayley graph spectral sparsifiers, along with an efficient algorithm for finding them. Our algorithm even extends to \emph{directed} Cayley graphs, if we instead ask only for cut sparsification instead of spectral sparsification.

We additionally study the sparsification of linear equations over non-abelian groups. In contrast to the abelian case, we show that for non-abelian valued equations, super-polynomially many linear equations must be preserved in order to approximately preserve the number of satisfied equations for any input. Together with our Cayley graph sparsification result, this provides a formal separation between Cayley graph sparsification and sparsifying linear equations.

%% file: content/intro.tex
\section{Introduction}

Graph sparsification is a popular technique pioneered by Bencz\'ur and Karger~\cite{BK96}, which promises to drastically decrease the size of a graph while still ensuring that the resulting graph has strong structural similarities to the original graph. In their landmark result,~\cite{BK96} showed that for any choice of $\eps \in (0,1)$, any graph $G = (V, E, w)$\footnote{We will be working with graphs with \emph{weighted} edges, where the weights are given by the function $w : E \rightarrow \mathbb{R}_{\ge 0}$.} can be sparsified to a graph $\wt{G} = (V, \wt{E}, \wt{w})$ such that $|\wt{E}| = \widetilde{O}(|V| / \eps^2)$, and that for every cut $T \subseteq V$, the size of the cut $T$ in $\wt{G}$ is within a $(1 \pm \eps)$ factor of the size of the cut $T$ in $G$.\footnote{The size of a cut $T$ is the sum of all the weights of the edges between $T$ and $V \backslash T$.} Because $|E|$ could have potentially been as large as $\Omega(|V|^2)$ initially, the work of \cite{BK96} provides a potential near-quadratic savings in the number of edges in the graph, while still ensuring that all exponentially many cuts in the graph are approximately preserved.

Because of this, cut sparsification and its later generalizations have found applications both as techniques for speeding up static algorithms \cite{BK96, spielman2004nearly, spielman2008graph} and for decreasing the space usage of sublinear algorithms \cite{AGM12b, KLMMS14, mcgregor2014graph, abraham2016fully}. As a consequence, sparsification has developed into a diverse field, with works studying hypergraph sparsification \cite{KK15, CKN20, KKTY21a, JLS22, Lee23}, constraint satisfaction problem (CSP) sparsification \cite{KK15, BZ20, KPS24, BrakensiekG}, and notions of matroid sparsification \cite{Qua23}. 

More recently, one prong of sparsification research has focused on \emph{Cayley graphs}, where sparsification has the potential to yield much stronger space savings than in conventional graph sparsification. Recall that a Cayley graph $H = \Cay(G, S, w)$ is specified by a group $G$, and a subset of (weighted) generators $S \subseteq G$, where the weights are given by $w : S \rightarrow \mathbb{R}_{\ge 0}$. The vertex set of $H$ is the group $G$, and two vertices $u$ and $v$ have an edge if and only if $u^{-1}v \in S$ (and the corresponding weight of the edge is simply the weight of the generator $u^{-1}v$). While graph sparsification techniques like those of \cite{BK96} can be arbitrarily applied to Cayley graphs, thereby yielding graph sparsifiers, there is no guarantee (and in fact it is exceedingly unlikely) that the resulting sparsifier will still maintain the \emph{Cayley graph} structure. While this may seem inconsequential, this actually leads to a fundamental inefficiency in the representation size of the graph: a Cayley graph is specified completely by its generating set, and so allows for a representation size of $\widetilde{\Theta}(|S|)$ bits, whereas an arbitrary graph is described by its edges, and so requires $\widetilde{\Theta}(|E|)$ bits to describe. Any graph sparsifier must be connected,\footnote{When sparsifying a connected graph.} and so when the Cayley graph structure is removed, describing the sparsifier requires $\widetilde{\Theta}(|G|)$ bits of space, which provides \emph{no improvement} over the original Cayley graph. Motivated by this, the work of Khanna, Putterman, and Sudan \cite{KPS24} introduced the notion of a \emph{Cayley graph sparsifier}, where the sparsifier is required to maintain the Cayley graph structure:

\begin{definition} 
    Let $G$ be a group, let $H = \mathrm{Cay}(G, S, w)$ be a Cayley graph over $G$, and let $\eps \in (0,1)$. We say that $\wt{H} = \mathrm{Cay}(G, \wt{S}, \wt{w})$ is a $(1 \pm \eps)$-Cayley graph cut sparsifier of $H$ if for every $T \subseteq V$, we have that 
    \[
    \wt{w}(\delta_{\wt{H}}(T)) \in (1 \pm \eps) \cdot w(\delta_{H}(T)),
    \]
    where $\wt{w}(\delta_{\wt{H}}(T))$ (resp. $w(\delta_H(T))$) denotes the size of the cut $T$ in the graph $\wt{H}$ (resp. $H$) weighted by $\wt{w}$ (resp. $w$).
\end{definition}

In fact, the work of \cite{KPS24} even studies a stronger notion of sparsification called \emph{spectral sparsification}:

\begin{definition}
    Let $G$ be a group, let $H = \mathrm{Cay}(G, S, w)$ be a Cayley graph over $G$, and let $\eps \in (0,1)$. We say that $\wt{H} = \mathrm{Cay}(G, \wt{S}, \wt{w})$ is a $(1 \pm \eps)$ Cayley graph spectral sparsifier of $H$ if for every $x \in \R^V$, we have that 
    \[
    \sum_{\wt{e} = (u,v) \in \wt{H}} \wt{w}_{\wt{e}} \cdot (x_u - x_v)^2 \in (1 \pm \eps) \cdot \sum_{e = (u,v) \in H} w_e \cdot (x_u - x_v)^2 \,.
    \]
    Note that when $x$ is restricted to $\{ 0, 1 \}^V$ instead of $\mathbb{R}^V$, this recovers the notion of cut sparsification.
\end{definition}

Spectral sparsification has seen extensive study in its own right, with several works  studying the connections between electrical networks and spectral sparsification \cite{ST11, spielman2008graph}, and showing how spectral sparsification opens the door for better sparsification algorithms (for instance, making such algorithms deterministic \cite{BSS09}).

In the context of Cayley graph sparsification, it was shown \cite{KPS24} that Cayley graphs over $G = \F_2^n$ admit $(1 \pm \eps)$ Cayley graph spectral sparsifiers which preserve only $\widetilde{O}(n / \eps^2)$ re-weighted generators, i.e., $|\wt{S}| = \widetilde{O}(n / \eps^2)$. In this sense, the description of the sparsifier requires only $\widetilde{O}(n^2 / \eps^2)$ bits, which is far less than any ordinary sparsifier, which requires $|G| = 2^n$ bits to describe. However, the existence of these Cayley graph sparsifiers uses the so-called ``code sparsification'' technique of \cite{KPS24}, which crucially relies on a connection between codeword weights in linear codes over $\F_2^n$ and eigenvalues in Cayley graphs over $\F_2^n$ that breaks down over larger fields. Thus, a key open question from their work, and indeed the main question we focus on is:

\begin{quote}
    \emph{Do all Cayley graphs (over any group) admit small Cayley graph sparsifiers?}
\end{quote}

\subsection{Our Results}

As our primary result, we answer the above question in the affirmative, showing that all Cayley graphs admit sparsifiers which preserve only poly-logarithmically many generators:

\begin{theorem}
\label{thm:sparsify-main}   
    Let $G$ be a finite group, let $H = \mathrm{Cay}(G, S, w)$ be an undirected Cayley graph over $G$ with polynomially bounded weights, and let $\eps \in (0,1)$. Then, there exists a re-weighted symmetric subset of generators $\wt{S} \subseteq S$ with weights $\wt{w} : \wt{S} \rightarrow \mathbb{R}_{\ge 0}$ such that $|\wt{S}| = O( \log^5(|G|) / \eps^2)$ and $\wt{H} = \mathrm{Cay}(G, \wt{S}, \wt{w})$ is a $(1 \pm \eps)$ spectral sparsifier of $H$. Furthermore, there is a randomized algorithm that computes $\wt{S}, \wt{w}$ in $\mathrm{poly}(|G|)$ time.
\end{theorem}

\begin{remark}[Extension to Schreier graphs]
\label{rem:schreier-main}
    Our proof of \Cref{thm:sparsify-main} works almost identically for Schreier graphs.
    More specifically, let $G$ be a group acting on a set $X$, any Schreier graph $\mathrm{Sch}(G, X, S)$ has a $(1\pm \eps)$ spectral sparsifier with $O(\log^5(|G| \cdot |X|) / \eps^2)$ re-weighted generators.
    See \Cref{rem:schreier-proof} for more details.
\end{remark}

Note that the sparsifiers from the above theorem can be described using only $O( \log^6(|G|) / \eps^2)$ bits of space, and thus provide a potential exponential savings in the bit complexity as compared to the starting Cayley graph. Further, while the above theorem only characterizes \emph{spectral} sparsifiability, it turns out that if we relax the spectral condition and instead require only \emph{cut} sparsification, the above result can be extended to arbitrary \emph{directed} Cayley graphs:

\begin{corollary} \label{cor:directed-cayley-sparsification}
    Let $G$ be a finite group, let $H = \mathrm{Cay}(G, S, w)$ be a (potentially directed) Cayley graph over $G$, with polynomially bounded weights, and let $\eps \in (0,1)$. Then, there exists a re-weighted subset of generators $\wt{S} \subseteq S$ with weights $\wt{w} : \wt{S} \rightarrow \mathbb{R}_{\ge 0}$ such that $|\wt{S}| = O( \log^5(|G|) / \eps^2)$ and $\wt{H} = \mathrm{Cay}(G, \wt{S}, \wt{w})$ is a $(1 \pm \eps)$ cut sparsifier of $H$. Furthermore, there is a randomized algorithm that computes $\wt{S}, \wt{w}$ in $\mathrm{poly}(|G|)$ time.
\end{corollary}

Finally, as mentioned before, Cayley graph sparsification over $\F_2^n$ has a strong connection (in fact, it is equivalent) to the so-called \emph{code sparsification} over $\F_2^n$.  Recently, code sparsification has seen generalizations to ``group-valued'' codes over $G^n$ \cite{khanna2025efficient}, where it was shown that all abelian groups admit sparsifiers of small size. Formally, in this setting we are given a group $G$, as well as a set of $m$ linear equations $C_1, \dots C_m$, where each linear equation operates on a subset of $n$ variables $x_1, \dots x_n \in \{0,1\}$. Each linear equation can be written as 
\[
C_j(x) = a_{j, 1}^{x_1} a_{j, 2}^{x_2} \dots a_{j, n}^{x_n},
\]
where $x_i \in \zo$, and each $a_{j, i} \in G$. 

The goal in this setting is to design a \emph{code-sparsifier} of the linear equations, namely, to find a set $S \subseteq [m]$, along with weights $\{w_i\}_{i \in S}$ such that \emph{for every} $x \in \zo^n$, 
\[
 \sum_{i \in S} w_i \cdot \mathbf{1}[C_j(x) \neq 1] \in (1 \pm \eps) \cdot \sum_{i \in [m]} \mathbf{1}[C_j(x) \neq 1],
\]
while also maintaining the set $S$ to be as small as possible. In the case when $G$ is an abelian group, it is known \cite{khanna2025efficient} that such sets $S$ exist of size $\widetilde{O}(n \mathrm{polylog}(|G|)/ \eps^2)$ (this should be thought of as being \emph{poly-logarithmic} in the group size, as this is implicitly over $G^n$).

However, for non-abelian groups, there is no known upper bound on sparsifier size. Thus, it is tempting to ask whether improvements in the sparsifiability of Cayley graphs over $G^n$ \emph{also} lead to improvements in code sparsification over $G^n$ (with a particular focus on non-abelian groups). In this direction, we prove our final result, which is a strong separation between these notions of sparsification:

\begin{theorem}[See \Cref{thm:lower-bound}] \label{thm:lower-bound-intro}
	There is a (non-abelian) group $G$, $|G| = O(1)$, along with a set $C$ of $G$-valued linear equations over $n$ variables such that for any $\eps \in (0,1)$, any $(1 \pm \eps)$ code sparsifier of $C$ must be of size $n^{\omega(1)} = \log^{\omega(1)}(|G^n|)$.
\end{theorem}

That is to say, in general non-abelian codes \emph{do not} admit poly-logarithmic size code sparsifiers, in stark contrast to their Cayley graph counterparts.

In the following section, we briefly highlight some of the techniques that go into our results before providing formal proofs in \Cref{sec:cayley-sparsification,sec:sparsification-lb}.

\parhead{Concurrent work.}
In a concurrent and independent work, Basu, Kothari, Liu, and Meka also proved \Cref{thm:sparsify-main} with similar parameters.

%% file: content/tech-overview.tex
\subsection{Technical Overview}
\label{sec:tech-overview}

Let $G$ be a group, and let $S$ be any set of generators closed under inverses.
Recall that our goal is to find a subset $\wt{S}\subseteq S$ along with a weight function $\wt{w}$ such that the weighted Cayley graph $\wt{H} = \Cay( G, \wt{S}, \wt{w} )$ is a spectral sparsifier of $H = \Cay(G, S)$ up to some desired quality $\eps$, and $|\wt{S}| \le \polylog |G| / \eps^2$.
The overall framework we use is one that is frequently employed in sparsification:
each $s\in S$ is assigned a probability $p_s\in[0,1]$;
then, the sparsifier is instantiated by prescribing to each $s$ weight $\frac{1}{p_s}$  and sampling $s$ with probability $p_s$.

Note that for $\wt{S}$ to be symmetric, we need to sample $s$ and $s^{-1}$ together whenever $s \neq s^{-1}$.
Thus, it is convenient to define $\ol{S} \subseteq S$ as follows: for each pair $s,s^{-1} \in S$ with $s \neq s^{-1}$, include exactly one of the two in $\ol{S}$; if $s = s^{-1}$, then include $s$ in $\ol{S}$.
Then, we will construct $\wt{S}$ by sampling generators from $\ol{S}$ and then include their inverses.

In our setting, we choose $p_s$ proportional to the \emph{importance} of $s$.
To motivate importance sampling, we discuss an example.
Consider the group \(G = \mathbb{F}_2^d\) with generating set
\(
S = \{e_1\} \;\cup\; \{\,v \in \mathbb{F}_2^d : v_1 = 0\}.
\)
Observe that since \(S\) generates \(G\), the Cayley graph \(H = \Cay(G,S)\) is connected and therefore has a strictly positive spectral gap. 
Observe that any subset \(\wt{S} \subseteq S\) that omits the distinguished generator \(e_1\) disconnects \(H\), forcing its spectral gap to drop to zero.
Hence, such a \(\wt{S}\) cannot spectrally sparsify \(S\).
In particular, it is imperative that any spectral sparsifier includes \(e_1\), and accordingly we say that \(e_1\) has \emph{importance} 1.

The notion of importance we use is derived from effective resistance sampling introduced by Spielman and Srivastava \cite{spielman2008graph} in the context of spectral sparsification of graphs.
The effective resistance of an edge $e$ is defined as $\|L_H^{\dagger/2} L_e L_H^{\dagger/2}\|_{\mathrm{op}}$, where $L_e$ is the Laplacian matrix of the single edge $e$.
Similarly, for $s \in S$, we define
\begin{equation} \label{eq:L-s}
    L_s = \begin{cases}
        2I - (A_s + A_{s^{-1}}) & s \neq s^{-1}, \\
        I - A_s & s= s^{-1}. \\
    \end{cases}
\end{equation}
where $A_s + A_{s^{-1}}$ is the adjacency matrix of the subgraph generated by $s, s^{-1}$.
Then, we define the importance of a generator $s \in S$ as
\[
    \imp(s) \coloneqq \norm*{L_H^{\dagger/2} L_s L_H^{\dagger/2}}_{\mathrm{op}} = \max_{v\in\bbS^{n-1}} \frac{v^{\top} L_s v}{v^{\top} L_H v},
\]
from which we define $p_s \coloneqq \min\braces*{ \imp(s)\cdot\frac{\log |G|}{\eps^2}, 1 }$.
Here, note that $L_s = L_{s^{-1}}$ by definition, hence $\imp(s) = \imp(s^{-1})$, and recall that we are sampling $s$ from $\ol{S}$ and including their inverses later.

The key challenge is in establishing that $\wt{S}$ is small with high probability. To this end, we prove that the number of generators with high importance (i.e., are kept with high probability) is small. Precisely, we show that for any $\alpha > 0$, the number of generators $s$ in $S$ such that $\imp(s) \ge \alpha$ is bounded by $\frac{\log^3 |G|}{\alpha}$ (\Cref{clm:boundImportanceNumber}). We outline our proof of this bound in the next section.

\subsubsection{Bounding the Number of Important Generators}

It is convenient for us to define an associated \emph{score function} of a generator $s \in \ol{S}$ where for $v\in\R^{|G|}$,
\[
    \score(s,v) \coloneqq \frac{v^{\top}L_s v}{v^{\top}L_H v}\mper
\]
We say a collection of generators $s_1,\dots,s_{\ell} \in \ol{S}$ and vectors $v_1,\dots,v_{\ell} \in \R^{|G|}$ are in \emph{upper triangular form} if $\score(s_i,v_j) \le \frac{\score(s_i,v_i)}{\log^2 |G|}$ for $j < i$.

Our proof of the bound on the number of important generators has two pieces:
\begin{enumerate}
    \item Proving that any set $T$ of generators with importance exceeding $\alpha$ contains a subset of generators in ``upper triangular position'' of size at least $\frac{\alpha}{\log^2 |G|}|T|$.
    \item Proving a bound of $O\parens*{\log |G|}$ on the number of important generators in upper triangular position.
\end{enumerate}
Combining the above two statements results in a bound of $\frac{\log^3 |G|}{\alpha}$ on $|T|$.

\parhead{Reduction to upper triangular position.}
We discuss how to find a large subset of important generators that are in upper triangular position.
An elementary observation is that for any $v$,
\[
    \sum_{s\in \ol{S}} L_s = L_H \mcom \quad
    \sum_{s\in \ol{S}} \score(s,v) = 1\mper  \numberthis \label{eq:score-at-most-1}
\]

Given generators $s_1,\dots,s_{\ell}$ with importance exceeding $\alpha$, 
we can find a large sequence of generators in upper triangular position by greedily picking generators (and their corresponding vector on which they have high score) that have low score on all previously chosen vectors. Notice that each chosen generator rules out at most $\frac{\log^2 |G|}{\alpha}$ other generators due to \Cref{eq:score-at-most-1}, so this greedy process runs for at least $\frac{\alpha}{\log^2 |G|} |T|$ steps.

\parhead{Bounding number of generators in upper triangular form.}
To prove a bound on the number of important generators in upper triangular position, we prove that for any pair of distinct collections of indices $1\le i_1 < \dots < i_{a}\le \ell$ and $1\le j_1 < \dots < j_b \le \ell$ where $i_1 < j_1$, we have $s_{i_1}\cdots s_{i_a} \ne s_{j_1}\cdots s_{j_b}$.
Thus, for every subset of $[\ell]$, we obtain a distinct group element, which implies $2^\ell \le |G|$ and consequently $\ell \le \log |G|$ as desired.

For the overview, it is instructive to consider the special case where $\score(s_i, v_j) = 0$ for $i \neq j$, and assume that $i_1 < j_1$.
Then, observe from \Cref{eq:L-s} that $L_s = (I- A_s)^{\top} (I - A_s)$ if $s \neq s^{-1}$ (with a factor of $1/2$ if $s = s^{-1}$), and thus $v_{i_1}^\top L_{s_j} v_{i_1} = 0$ implies that $\|v_{i_1} - A_{s_j} v_{i_1}\| = 0$.
That is, the operator $A_{s_j}$ does not change $v_{i_1}$ for all $j \neq i_1$.
In particular, this means that $A_{s_{j_1} s_{j_2} \cdots s_{j_b}} v_{i_1} = v_{i_1}$.
On the other hand, since $\score(s_{i_1}, v_{i_1}) > 0$, we must have $A_{s_{i_1} s_{i_2} \cdots s_{i_a}} v_{i_1} \neq v_{i_1}$.
This establishes that $s_{i_1}\cdots s_{i_a} \ne s_{j_1}\cdots s_{j_b}$.

More generally, we prove a quantitative version of the above heuristic using the guarantee of the upper triangular form.
We prove a lower bound on $\norm*{(A_{s_{i_1} \cdots s_{i_a}} - A_{s_{j_1}\cdots s_{j_b}})v_{i_1}}$, which implies $s_{i_1}\cdots s_{i_a} \ne s_{j_1}\cdots s_{j_b}$.
See the proof of \Cref{clm:boundImportanceNumber} for more details.

%% file: content/sparsification.tex
\section{Cayley Graph Sparsification} \label{sec:cayley-sparsification}

To start, we present our argument for deriving spectral sparsifiers for undirected Cayley graphs over arbitrary non-abelian groups.
We first focus on unweighted Cayley graphs (\Cref{thm:nonabelianSparsifier}).
Then, we generalize the result to weighted Cayley graphs in \Cref{sec:sparsify-weighted-cayley}  (\Cref{cor:sparsify-weighted}), and further to directed Cayley graphs in \Cref{sec:sparsify-directed-cayley} (\Cref{thm:sparsify-directed}).

Unlike those over abelian groups, non-abelian Cayley graphs do not necessarily have a canonical set of eigenvectors. Thus, instead of analyzing the action of the generating set to the canonical eigenvectors as in previous work \cite{KPS24}, our spectral sparsifiers in the non-abelian setting will require analyzing the contribution of generators to \emph{all vectors}.

Given an undirected Cayley graph $H$ on a group $G$ with generating set $S$, we define the \emph{score function}:
\[
    \score(s, v) \coloneqq  \frac{v^{\top} L_s v}{v^{\top} L_H v} \mcom
\]
where $L_s$ is the Laplacian matrix corresponding to the generator $s$ and its inverse as in \Cref{eq:L-s}.

We can then define importances in the following manner:
\begin{definition}[Importance]
    For a non-abelian Cayley graph $H$ with generating set $S \subseteq G$, for a generator $s \in S$, we define the importance of the generator $s$ as
    \[
    \mathrm{imp}(s) = \max_{v \in \mathbb{S}^{n-1}} \score(s, v).
    \]
\end{definition}
As discussed in \Cref{sec:tech-overview}, if we replace $L_s$ with $L_e$ for an edge $e \in E$, then the above importance definition is exactly the effective resistance of $e$.

\begin{observation}[Efficiently computing the importance]
\label{obs:computing-importance}
    Using a change of variables $y = L_H^{\dagger/2}v$,
    \begin{align*}
        \imp(s) = \max_{y\in\R^n} \frac{y^{\top} L_{H}^{\dagger/2} L_s L_{H}^{\dagger/2} y}{y^{\top} y}
        = \norm*{L_H^{\dagger/2} L_s L_H^{\dagger/2}}_{\mathrm{op}} ,
    \end{align*}
    which reduces to an eigenvalue computation of $L_{H}^{\dagger/2} L_s L_{H}^{\dagger/2}$, which can be done in polynomial time.
\end{observation}

The following is the randomized algorithm to sparsify Cayley graphs, which runs in polynomial time.
It is essentially an importance sampling algorithm based on the importance of each generator.

\vspace{1em}
\noindent\rule{\textwidth}{0.4pt}
\vspace{-1em}
\begin{algorithm}[Cayley graph sparsification]
\label{alg}
    Given a Cayley graph $H$ over a group $G$ with generating set $S \subseteq G$.
    \begin{enumerate}
        \item Compute $\imp(s)$ for each $s\in \ol{S}$.
        \item Set probabilities $p_s = \min\left\{C \cdot \frac{\imp(s) \log|G|}{\eps^2}, 1\right\}$ for a large enough constant $C$, and assign weight $1/p_s$ to $s$.
        \item For each $s\in \ol{S}$, sample $s$ (and add in $s^{-1}$ if $s \neq s^{-1}$) with probability $p_s$.
    \end{enumerate}
\noindent\rule{\textwidth}{0.4pt}
\end{algorithm}

To prove \Cref{thm:sparsify-main}, we must show that the number of generators sampled by \Cref{alg} is at most $\polylog(|G|)$ (with high probability).
Since the sampling probability is proportional to the importance, we first need an upper bound on the number of generators with ``large'' importance.

\subsection{Bounding Important Generators}
\label{sec:bounding-important-generators}

We now state the key technical lemma we use to analyze the sparsification procedure.
\begin{lemma}\label{clm:boundImportanceNumber}
     Let $G$ be a group, $S\subseteq G$ a generating set, and $H = \Cay(G, S)$.
     For any $\alpha < 1$, the number of generators $s \in S$ with $\mathrm{imp}(s) > \alpha$ is at most $O(\frac{\log^3|G|}{\alpha})$.
\end{lemma}

\begin{proof}
    Let $S_{\alpha}$ denote the generators $s$ with $\imp(s) \ge \alpha$.
    We first prove that we can find a large ordered collection of elements $s_1,\dots,s_{\ell} \in S_{\alpha}$, along with vectors $v_1,\dots,v_{\ell}$ that are in \emph{upper-triangular form}, i.e.:
    \begin{enumerate}
        \item $\score(s_i,v_i)\geq \alpha$ for all $i\in[\ell]$.
        \item $\score(s_i,v_j)\leq \frac{\alpha}{C\log^2|G|}$ for $i > j$, where $C$ is a large constant.
        \item $\ell \ge \frac{\alpha}{C \log^2|G|} \abs*{S_{\alpha}}$.
    \end{enumerate}
    We will show that any collection of generators and vectors in upper triangular form must have size at most $\log_2|G|$.
    Note that this then implies the desired statement.

    \parhead{Reduction to upper-triangular form.}
    To prove this, we construct $s_1,\dots,s_{\ell}$ along with $v_1,\dots,v_{\ell}$ via an iterative process.
    We initialize the process with all elements of $S_{\alpha}$ labeled \texttt{permitted}, and a counter $i$ to $1$, and then perform the following procedure, while $S_{\alpha}$ still contains a \texttt{permitted} generator.
    \begin{enumerate}
        \item Pick a \texttt{permitted} generator $s\in S_{\alpha}$, and let $s_i = s$.
        Choose unit $v_i\in\R^{G}$ such that $\score(s_i,v_i)\ge\alpha$, which always exists by definition of importance.
        \item For every $s\in S_{\alpha}$ such that $\score(s,v_i) \ge \frac{\alpha}{C\log^2 |G|}$, change the label of $s$ to \texttt{forbidden}.
        \item Increment $i$ by $1$.
    \end{enumerate}
    The first and second requirements of the upper-triangular form are satisfied by design.
    To prove the claimed lower bound on $\ell$, observe that for any vector $v$,
    \(
        \sum_{s\in \ol{S}} \score(s,v) = 1\mcom
    \)
    and hence, the number of generators whose label changes from \texttt{permitted} to \texttt{forbidden} in a single iteration is at most $\frac{C\log^2|G|}{\alpha}$.

    Consequently, the procedure lasts for $\frac{\alpha}{C\log^2|G|}\abs*{S_{\alpha}}$ iterations, and the claimed lower bound on $\ell$ follows.

    \parhead{Bound on number of generators in upper-triangular form.}
    Assume for contradiction that $\ell > \log |G|$, and define $\ell' = \ceil{\log |G|}$.
    For any subset $T\subseteq[\ell']$, equal to $\{i_1,\dots,i_t\}$ where $1\le i_1 < \dots < i_t\le \ell'$, we use $\prod_{i\in T} s_i$ to denote $s_{i_1}\cdots s_{i_t}$.
    We will prove that for any pair of subsets $T_1 \neq T_2 \subseteq [\ell']$, we have $\prod_{i \in T_1} s_i \neq \prod_{i \in T_2} s_i$.
    Note that this immediately implies that the group $G$ has at least $2^{\ell'} > |G|$ distinct elements, which is a contradiction.

    Now, for each group element $j$, let $A_j \in \zo^{n \times n}$ denote the adjacency matrix induced by the Cayley graph with only $j$ as a generator.
    Importantly, we have that $A_{a\cdot b} = A_a \cdot A_b$.
    For a subset $T\subseteq[\ell']$, we define 
    \(
        A_T = \prod_{i \in T} A_i\mper
    \)
    Assume for the sake of contradiction for some pair $T_1,T_2\subseteq[\ell']$ that $\prod_{i \in T_1} s_i = \prod_{i \in T_2} s_i$.
    This implies that 
    \[
        A_{T_1} = \prod_{j \in T_1} A_j = \prod_{j \in T_2} A_j = A_{T_2}.
    \]
    If $T_1, T_2$ share a prefix of elements, we can simply remove these elements (resp.\ their matrices) from both products. Thus, we may assume without loss of generality that the first element $i$ in $T_1$ is not in $T_2$.

    As a consequence of $A_{T_1} = A_{T_2}$, we get that $A_{T_1} v_i = A_{T_2} v_i$.
    To obtain a contradiction, we will use the fact that we have generators and vectors in upper triangular form to prove $\norm*{A_{T_1} v_i - v_i} > \norm*{A_{T_2}v_i - v_i}$.
    In service of this, we prove the following two inequalities.
    \begin{align}
        \norm*{ A_{s_i}(v_i + \Delta) - v_i } &\ge \sqrt{\alpha v_i^{\top}L_Hv_i} - \norm*{\Delta}   \label{eq:shift-vector-lot} \\
        \norm*{ A_{s_j}(v_i + \Delta) - v_i } &\le \frac{\sqrt{2\alpha v_i^{\top}L_H v_i}}{\sqrt{C}\log |G|} + \norm*{\Delta}\mcom   \label{eq:shift-vector-bound}
    \end{align}
    where $\Delta$ is any vector, and $j > i$.
    
    We prove \Cref{eq:shift-vector-lot} via the following chain of inequalities.
    \begin{align*}
        \norm*{A_{s_i}(v_i + \Delta) - v_i} &\ge \norm*{v_i - A_{s_i}v_i} - \norm*{A_{s_i}\Delta} \\
        &\ge \sqrt{ v_i^{\top} L_{s_i} v_i } - \norm*{\Delta} \\
        &\ge \sqrt{\alpha v_i^{\top}L_H v_i} - \norm*{\Delta}\mcom
    \end{align*}
    where the first step follows from the triangle inequality, the second uses that $A_{s_i}$ is an orthogonal matrix and $\|v_i - A_{s_i} v_i\|^2 = 2\|v_i\|^2 - 2 v_i^\top A_{s_i} v_i = (1 + \ind(s_i= s_i^{-1})) \cdot v^\top L_{s_i} v_i$ (recall the definition of $L_{s_i}$ in \Cref{eq:L-s}), and the final line follows from the upper triangular form.

    We prove \Cref{eq:shift-vector-bound} similarly below.
    \begin{align*}
        \norm*{A_{s_j}(v_i + \Delta) - v_i} &\le \norm*{v_i - A_{s_j}v_i} + \norm*{A_{s_i}\Delta} \\
        &\le \sqrt{2 v_i^{\top} L_{s_j} v_i} + \norm*{\Delta} \\
        &\le \frac{\sqrt{2\alpha v_i^{\top}L_H v_i}}{\sqrt{C}\log |G|} + \norm*{\Delta}\mper
    \end{align*}

    Since all elements of $T_1\setminus i$ and $T_2$ are greater than $i$, by an iterative application of \Cref{eq:shift-vector-bound}, we have:
    \begin{align}
        \norm*{A_{T_1\setminus i} v_i - v_i} &\le \frac{\sqrt{2\alpha v_i^{\top} L_H v_i}}{\sqrt{C}\log|G|} |T_1 \setminus i| \label{eq:T_1-without-i} \\
        \norm*{A_{T_2}v_i - v_i} &\le \frac{\sqrt{2\alpha v_i^{\top} L_H v_i}}{\sqrt{C}\log|G|} |T_2| \le \frac{2}{\sqrt{C}}\sqrt{\alpha v_i^{\top}L_H v_i} \mper \label{eq:T_2}
    \end{align}
    On the other hand, by \Cref{eq:shift-vector-lot} and \Cref{eq:T_1-without-i}, we have:
    \begin{align*}
        \norm*{ A_{T_1} v_i - v_i } &= \norm*{ A_{s_i} (v_i + A_{T_1\setminus i} v_i - v_i) - v_i } \\
        &\ge \sqrt{\alpha v_i^{\top} L_H v_i} - \norm*{ A_{T_1\setminus i} v_i - v_i } \\
        &\ge \sqrt{\alpha v_i^{\top} L_H v_i} - \frac{\sqrt{2\alpha v_i^{\top} L_H v_i}}{\sqrt{C}\log|G|} |T_1 \setminus i| \\
        &\ge \sqrt{\alpha v_i^{\top} L_H v_i} \parens*{ 1 - \frac{2}{\sqrt{C}} }    \numberthis \label{eq:T_1-shift}
    \end{align*}
    By choosing $C$ as a large enough constant, by \Cref{eq:T_2,eq:T_1-shift}, we have $\norm*{A_{T_1}v_i - v_i} > \norm*{A_{T_2}v_i - v_i}$, which leads to the desired contradiction starting from assuming $A_{T_1} = A_{T_2}$.
\end{proof}

\subsection{Spectral Sparsifiers for Undirected Cayley Graphs}

Now, it remains to show how we can use the bound on the number of important generators to create spectral sparsifiers. To do this, we first recall the matrix Chernoff bound:

\begin{theorem}[\cite{Tropp15}] \label{thm:matrixChernoff}
    Let $X = X_0 + \sum_{i = 1}^m X_i$, where $X_0$ is deterministic and positive semidefinite, and for $i \ge 1$, each $X_i$ is an independent, $n \times n$ random, positive semidefinite matrix, with $\Pr[\norm{X_i}_{\mathrm{op}} \leq R] = 1$.
    If we let $\mu_{\min} = \lambda_{\min}(\E[X])$, and $\mu_{\max} = \lambda_{\max}(\E[X])$, then for all $\eps \in (0,1)$, we have that 
    \begin{enumerate}
        \item $\Pr[\lambda_{\min}(X) \leq (1 - \eps) \mu_{\min}] \leq n \cdot e^{- \eps^2 \mu_{\min} / 2R}$.
        \item $\Pr[\lambda_{\max}(X) \geq (1 + \eps) \mu_{\max}] \leq n \cdot e^{- \eps^2 \mu_{\min} / 3R}$.
    \end{enumerate}
\end{theorem}

We will use this theorem to prove the following, which immediately implies the unweighted version of \Cref{thm:sparsify-main}:

\begin{theorem}\label{thm:nonabelianSparsifier}
    Let $H$ be an undirected, unweighted Cayley graph over a group $G$ with generating set $S \subseteq G$.
    Suppose $\widetilde{S} \subseteq S$ is the  generating set, with weights $\widetilde{w}: \widetilde{S} \to \R_{\geq 0}$, sampled by \Cref{alg}.
    Then, with probability $1-1/\mathrm{poly}(|G|)$, we have:
    \begin{enumerate}
        \item $|\widetilde{S}| \leq O\parens*{\log^4 |G| \cdot \log |S| / \eps^2}$.
        \item $\widetilde{H} = \mathrm{Cay}(G, \widetilde{S}, \widetilde{w})$ is a $(1 \pm \eps)$ spectral-sparsifier of $H = \mathrm{Cay}(G, S)$.
    \end{enumerate}
\end{theorem}

\begin{proof}
    We begin by proving the sparsity condition.
    Observe that:
    \[
        \frac{\eps^2}{C\log|G|} \E |\wt{S}| \le \int_0^1 \#\{s\in S: \imp(s) \ge \alpha\} \dif \alpha \le 1 + \int_{1/|S|}^1 \frac{\log^3 |G|}{\alpha} \le O\parens*{\log^3 |G|\cdot \log |S|}\mcom
    \]
    where the inequality uses \Cref{clm:boundImportanceNumber}.
    This translates to $\E|\wt{S}| \le O\parens*{ \frac{\log^4 |G|\,\cdot\,\log |S|}{\eps^2} }$, as desired.
    By the standard Chernoff bound, we have that $|\wt{S}|$ is upper bounded by the above with high probability.

    Thus, the remainder of this proof is dedicated to the proof that $\wt{H}$ is a spectral sparsifier of $H$ with high probability.
    For each generator $s \in S$, we define a random matrix $X_s$ as follows:
    \[
    X_s = \begin{cases}
        \frac{1}{p_s} \cdot L_{H}^{\dagger/2} L_{s}  L_{H}^{\dagger/2} \quad & \text{w.p. } p_s \\
        0  & \text{else.}
    \end{cases}
    \]
    Note then that we immediately have that 
    \[
    X = \sum_{s\in S: p_s=1} X_s + \sum_{s \in S: p_s < 1}X_s,
    \]
    and so 
    \[
    \E[X] = \sum_{s \in \ol{S}} L_{H}^{\dagger/2} L_{s}  L_{H}^{\dagger/2}  = L_{H}^{\dagger/2} L_H L_{H}^{\dagger/2} = I,
    \]
    where we are restricting our attention to the image of all the above matrices (namely the vectors orthogonal to $\mathbf{1}$).

    We will apply matrix Chernoff to the above with $X_0$ as the first term.
    It remains to bound the norm of each $X_s$ for $s$ such that $p_s < 1$.
    We have that 
    \[
    \Vert X_s \Vert = \frac{1}{p_s} \Vert L_{H}^{\dagger/2} L_{s} L_{H}^{\dagger/2} \Vert = \frac{1}{p_s} \cdot \max_{v \in \mathbb{S}^{|G|-1}} \frac{v^\top L_s  v}{v^\top L_H v} = \frac{1}{p_s} \cdot \mathrm{imp}(s) = \frac{\eps^2}{C \log|G|} = R. 
    \]
    Plugging this into the statement of \cref{thm:matrixChernoff}, we obtain that (for a sufficiently large choice of constant $C$):
    \begin{enumerate}
        \item $\Pr[\lambda_{\min}(X) \leq (1 - \eps) \mu_{\min}] \leq |G| \cdot e^{- \eps^2 \mu_{\min} / 2R} \leq n \cdot e^{- \eps^2 C \log|G| / 2\eps^2} \leq 1 / \mathrm{poly}(|G|)$.
        \item $\Pr[\lambda_{\max}(X) \leq (1 + \eps) \mu_{\max}] \leq |G| \cdot e^{- \eps^2 \mu_{\min} / 3R} \leq n \cdot e^{- C \eps^2 \log|G| / 3\eps^2}  \leq  1/ \mathrm{poly}(|G|)$.
    \end{enumerate}
    Thus, with probability $1 - 1 / \mathrm{poly}(|G|)$, we have that the resulting matrix $X$ satisfies
    \[
    (1 - \eps) I \preceq X \preceq (1 + \eps) I.
    \]
    Letting $\widetilde{S}$ denote the $s$ for which $X_s$ was sampled, we then see that 
    \[
    (1 - \eps) I \preceq \sum_{s \in \widetilde{S}} \frac{1}{p_s}L_{H}^{\dagger/2} L_{s} L_{H}^{\dagger/2}  \preceq (1 + \eps) I,
    \]
    which implies that 
    \[
    (1 - \eps) L_H \preceq \sum_{s \in \widetilde{S}} \frac{1} {p_s} \cdot L_s = L_{\widetilde{H}}\preceq (1 + \eps) L_H.
    \]
    This proves that $\widetilde{H}$ is indeed a $(1 \pm \eps)$ spectral-sparsifier of $H$.
\end{proof}

\begin{remark}  \label{rem:schreier-proof}
    As mentioned in \Cref{rem:schreier-main}, our proof of the existence of Cayley sparsifiers on few generators naturally extends to Schreier graphs.
    Indeed, observe that the only property we made use of in the Cayley graph setting was that the adjacency matrices of group elements $a,b$ and $a\cdot b$ satisfy $A_a\cdot A_b = A_{a\cdot b}$, which is a relation that also holds in Schreier graphs.
\end{remark}

\begin{remark}
    In the setting of graph sparsification, \cite{BSS09} improved upon \cite{BK96,spielman2008graph} to give \textit{constant degree} sparsifiers for any graph. However, in our setting, it is clear to see that this is \emph{not} possible. I.e., there are abelian groups $G$ (like $\mathbb{F}_2^n$) which require $\Omega(\log |G|)$ generators to be connected, and more generally, abelian groups require $\Omega\left(\log|G|\right)$ generators to have diameter $O(\log |G|)$ (a necessary condition for expansion). Therefore, any sparsifier for $S=G$ (i.e. the complete graph) requires $\Omega(\log |G|)$ generators.

    A second interpretation of \cite{BSS09} is that it shows that sparsifying general graphs is no harder than sparsifying the complete graph. However, again for Cayley graph sparsification, this does not turn out to be true in full generality. Indeed, we can consider any non-abelian group $G$ which admits expanding Cayley graphs of constant degree, yet has an abelian subgroup $A\subset G$ of size $\mathrm{poly}(|G|)$. If we consider the set of generators $S=A$, then sparsifying the Cayley graph requires sparsifying the sub-group, and thus requires $\Omega(\log |A|)$ generators to sparsify as this is exactly the setting of abelian Cayley graph sparsification. On the other hand, sparsifying the complete graph (i.e. $S=G$) would admit sparsifiers of constant-degree. 

    However, there are still many open questions for Cayley graph sparsification. Perhaps most centrally, can we improve the sparsifiers to only preserve $O(\log |G| / \eps^2) $ re-weighted generators (as opposed to polylogarithmic)? 
\end{remark}

\subsection{Generalizing to Weighted Cayley Graphs}
\label{sec:sparsify-weighted-cayley}

As a corollary to \Cref{thm:nonabelianSparsifier}, we can also obtain the following \emph{weighted} Cayley graph sparsification result, which implies \Cref{thm:sparsify-main}:

\begin{corollary} \label{cor:sparsify-weighted}
    Let $H = \mathrm{Cay}(G, S, w)$ be a weighted Cayley graph, with every weight at least $1$, and let $w_{\max}$ denote the max weight. Then, there is an efficient algorithm finding a $(1 \pm \eps)$ Cayley graph sparsifier which preserves \[
    O\left(\log^4 |G| \log\left ( \frac{|S| \cdot w_{\max}}{\eps} \right ) / \eps^2 \right)
    \]re-weighted generators with high probability. 
\end{corollary}

\begin{proof}
    For every generator $s$ of weight $w_s$, we replace $s$ with $\lfloor 10 w_s / \eps  \rfloor$ unweighted copies of $s$. Let us denote this Cayley graph by $H_{\mathrm{unweighted}}$. We claim that for every $v \in \R^n$, 
    \[
    v^{\top} \left (\frac{\eps}{10} \cdot L_{H_{\mathrm{unweighted}}}\right ) v \in (1 \pm \eps/10) v^{\top} L_H  v.
    \]
    To see why, for each fixed generator $s$, we have that 
    \[
    (1 - \eps / 10) w_s \leq \frac{\eps}{10} \cdot \lfloor 10 w_s / \eps  \rfloor \leq w_s,
    \]
    and so 
    \[
       (1 - \eps / 10) w_s \cdot  v^{\top} L_s  v \leq \frac{\eps}{10} \cdot \lfloor 10 w_s / \eps  \rfloor \cdot  v^{\top} L_s  v \leq w_s \cdot  v^{\top} L_s  v.
    \]
    Summing across the generators then yields the stated claim.

    Finally, $H_{\mathrm{unweighted}}$ has only unweighted generators, and so by \cref{thm:nonabelianSparsifier}, we can find a set $\widetilde{S}$ with weights $\widetilde{w}$ such that $\mathrm{Cay}(G, \widetilde{S}, \widetilde{w})$ is a $(1 \pm \eps/10)$ spectral sparsifier of $H_{\mathrm{unweighted}}$. By composing accuracy, we then have that $\mathrm{Cay}(G, \widetilde{S}, \frac{\eps}{10} \cdot \widetilde{w})$ is a $(1 \pm \eps/10)^2 \in (1 \pm \eps)$ spectral sparsifier of $H$. Further, because $H_{\mathrm{unweighted}}$ only has $O(|S| w_{\max}/\eps)$ unweighted generators, $\widetilde{S}$ is of size $O\left(\log^4|G| \log\left ( \frac{|S| \cdot w_{\max}}{\eps}\right )/\eps^2 \right)$, as we desire.
\end{proof}

\subsection{Cut Sparsification for Directed Cayley Graphs}
\label{sec:sparsify-directed-cayley}

In this section, we show how to \emph{relax} the undirected-ness condition when we instead only require cut sparsifiers of Cayley graphs. To do so, we first use the following simple observation:

\begin{claim}
    Let $H = \mathrm{Cay}(G, \{t\})$ be a Cayley graph with a single generator $t$ where $t \neq t^{-1}$, and let $U(H) = \mathrm{Cay}(G, \{t, t^{-1} \})$ be the undirected version of $H$. Then, for any $T \subseteq G$,
    \[
    w(\delta_H(T)) = \frac{w(\delta_{U(H)}(T))}{2}.
    \]
\end{claim}

\begin{proof}
    Observe that for any single generator $t$ with $t \neq t^{-1}$, each connected component of the Cayley graph $\mathrm{Cay}(G, \{t\})$ is Eulerian because every vertex has in-degree $1$ and out-degree $1$. 

    Now, a well-known fact (see for instance \cite{DBLP:conf/icalp/CenCP021}) is that Eulerian graphs have perfect cut-balance, i.e., that 
    \[
    w(\delta_H(T)) = w(\delta_H(\ol{T})).
    \]
    Finally, we claim that $w(\delta_H(\ol{T})) = w_{\mathrm{Cay}(G, \{t^{-1}\})}(T)$. This is because an edge $(u,v)$ is in $H$ if and only if $(v,u)$ is in $\mathrm{Cay}(G, \{t^{-1}\})$. To, for every edge $(u,v)$ crossing from $\ol{T}$ to $T$ in $H$, there is exactly the edge $(v, u)$ crossing from $T$ to $\ol{T}$ in $\mathrm{Cay}(G, \{t^{-1}\})$. All together, we see that 
    \[
    w(\delta_H(T)) = w(\delta_H(\ol{T})) = w(\delta_{\mathrm{Cay}(G, \{t^{-1}\})}(T)),
    \]
    and so 
    \[
    w(\delta_H(T)) = \frac{w(\delta_H(T)) + w(\delta_{\mathrm{Cay}(G, \{t^{-1}\})}(T))}{2} = \frac{w(\delta_{U(H)}(T))}{2}.
    \]
\end{proof}

Below, we will use $U(H)$ to denote the undirectification of the entire Cayley graph, i.e., including $s^{-1}$ for every generator $s \in S$.

\begin{theorem} \label{thm:sparsify-directed}
    Let $H$ be an arbitrary (potentially directed) Cayley graph over a non-abelian group $G$ with generating set $S$. Then, there is a polynomial time algorithm which recovers a weighted subset of generators $\widetilde{S}$ such that (with high probability):
    \begin{enumerate}
        \item $|\widetilde{S}| \leq O(\log^4 |G| \log(|S| / \eps)/ \eps^2)$.
        \item With high probability, $\widetilde{H} = \mathrm{Cay}(G, \widetilde{S})$ is a $(1 \pm \eps)$ cut-sparsifier of $H = \mathrm{Cay}(G, S)$.
    \end{enumerate}
\end{theorem}

\begin{proof}
    For generators $s \in S$ with $s = s^{-1}$, we may simply treat them as undirected edges and sparsify via \Cref{thm:nonabelianSparsifier}.
    Thus, we can just focus on generators with $s \neq s^{-1}$.
    
    Given the Cayley graph $H$, we create its undirected version $U(H)$ (i.e., including the generator $s^{-1}$ for every generator $s$), and plug it into \cref{thm:nonabelianSparsifier}, yielding a Cayley graph $\widetilde{U(H)}$, with generating set $\widetilde{U(S)}$. Note that the generators in $\widetilde{U(S)}$ come in pairs $(s, s^{-1})$ (if $s \neq s^{-1}$), where the generator $s$ was originally in $H$. 

    To retrieve our sparsifier $\widetilde{H}$, we simply take very pair $(s, s^{-1})$, and keep only the generator $s$. We have the property that for every $T \subseteq G$,
    \[
    w(\delta_{\widetilde{H}}(T)) = \frac{1}{2} \cdot w(\delta_{\widetilde{U(H)}}(T)) \in \frac{1}{2}(1\pm \eps)w(\delta_{U(H)}(T)) \in (1 \pm \eps) w(\delta_{H}(T)).
    \qedhere
    \]
\end{proof}

\section{Sparsification Lower Bounds for Linear Equations over Non-abelian Groups} \label{sec:sparsification-lb}

With the established upper bounds for sparsifying Cayley graphs over non-abelian groups, it is tempting to try to extend the above techniques to sparsify arbitrary ``linear'' equations over non-abelian groups. Recall that in this setting we are given a group $G$, as well as a set of $m$ linear equations $C_1, \dots C_m$, where each linear equation operates on a subset of $n$ variables $x_1, \dots x_n$.
Each linear equation can be written as 
\[
    C_j(x) = a_{j, 1}^{x_1} a_{j, 2}^{x_2} \dots a_{j, n}^{x_n},
\]
where $x_i \in \zo$, and each $a_{j, i} \in G$. 

The goal in this setting is to design a \emph{code-sparsifier} of the linear equations, namely, to find a set $S \subseteq [m]$, along with weights $\{w_i\}_{i \in S}$ such that \emph{for every} $x \in \zo^n$, 
\[
 \sum_{i \in S} w_i \cdot \mathbf{1}[C_j(x) \neq 1] \in (1 \pm \eps) \cdot \sum_{i \in [m]} \mathbf{1}[C_j(x) \neq 1],
\]
while also maintaining the set $S$ to be as small as possible. In the case when $G$ is an abelian group, it is known \cite{khanna2025efficient}, that such sets $S$ exist of size $\widetilde{O}(n / \eps^2)$ (this should be thought of as being \emph{logarithmic} in the group size, as this is implicitly over $G^n$). Ultimately, we will show that, despite the fact that $\mathrm{polylog}$-size sparsifiers exist for \emph{Cayley graphs} over $G^n$, the same does not hold true for these linear equations over non-abelian $G^n$.

Ultimately, our goal is to invoke a result from \cite{khanna2025efficient} on lower bounds for sparsifying \emph{CSPs}.
In this setting, we are given a function $P: \zo^r \rightarrow \zo$, and along with different \emph{constraints} $C_1, \dots C_m$ by applying $P$ to different subsets of $r$ variables. So, $C_1 = P(x|_{T_1}), \dots C_m = P(x|_{T_m})$, where $T_i$ is an arbitrary ordered subset of variables. As before, a sparsifier is a set $S \subseteq [m]$, along with weights $\{w_i\}_{i \in S}$ such that \emph{for every} $x \in \zo^n$,
\[
 \sum_{i \in S} w_i \cdot P(x|_{T_i}) \in (1 \pm \eps) \cdot \sum_{i \in [m]} P(x|_{T_i}).
\]

With this, the lower bound of \cite{khanna2025efficient} can be stated as:

\begin{theorem}[\cite{khanna2025efficient}, Theorem 3.4]\label{thm:andLB}
    If there is a function $P(x_1, \dots x_{\ell})$, $P: \zo^{\ell} \rightarrow \zo$, along with a restriction $\pi: \{x_1, \dots x_{\ell} \} \rightarrow \{y_1, \dots y_r, \neg y_1, \dots \neg y_r, 0, 1\}$ such that $P(\pi(x_1), \dots \pi(x_{\ell})) = \mathbf{AND}_{r}(y_1, \dots y_r)$, then there exist CSP instances on $n$ variables using the function $P$ such that any sparsifier requires preserving $\Omega((n / \ell)^{r})$ constraints. 
\end{theorem}

Note that in our setting, the function $P$ will be exactly this indicator of the linear equation being non-one, i.e., $P(x) = \mathbf{1}[C_j(x) \neq 1]$. Towards our lower bound, we show the following:

\begin{claim}\label{clm:simulateAND}
    For any constant $r > 0$, there is a linear equation $C(x_1, \dots x_{K_r})$ over $K_r$ variables $x_1, \dots, x_{K_r}$ with coefficients in $\mathbf{D}_3$ (the dihedral group with $6$ elements) along with a mapping $\pi: \{x_1, \dots x_{K_r}\} \rightarrow \{y_1, \dots y_r\}$ such that
    \[
    \mathbf{1}[C(\pi(x_1), \dots \pi(x_{K_r})) \neq 1] = \textbf{AND}_r(y_1, \dots y_r),
    \]
    and $K_r \leq 3^r$.
\end{claim}

\begin{proof}
    We first construct the linear equation over the $y$ variables, and then show how this can be represented with a linear equation over the $x$ variables along with a restriction map $\pi$. 

    To start, we observe that because $\mathbf{D}_3$ is a non-abelian group, there exist elements $b_1, b_2$ such that $b_1b_2 \neq b_2b_1$, or equivalently, such that $b_1b_2b_1^{-1}b_2^{-1} \neq 1$. So, for our base case, we show that this allows for a simple construction of $\textbf{AND}_2$. Indeed, we can consider the linear equation 
    \[
    L_2(y_1, y_2) = b_1^{y_1} b_2^{y_2} b_1^{-y_1} b_2^{-y_2}.
    \]
    By simple calculations, we can see that when $(y_1, y_2) = (0,0), (0,1), (1,0)$, the above expression $L_2(y_1, y_2) = 1$. On the other hand, when $(y_1, y_2) = (1,1)$, $L_2(y_1, y_2) = b_1b_2b_1^{-1}b_2^{-1} \neq 1$. In this manner, $L_2(y_1, y_2)$ simulates $\textbf{AND}_2$, as it evaluates to non-trivial group elements if and only if $y_1 = y_2 = 1$.

    Now, let us assume inductively that we have constructed $L_i(y_1, \dots y_i)$ such that for any $(y_1, \dots y_i) \neq (1, \dots 1)$, $L_i(y_1, \dots y_i) = 1$, and otherwise, $L_i(1, \dots 1) = z_i$, where $z_i \in \mathbf{D}_3$ is some non-identity group element. In particular, it is known that in $\mathbf{D}_3$, every non-identity element $z_i$ has another element $t_i$ such that $z_it_i \neq t_iz_i$. Notationally, we will refer to this element $t_i$ as $b_{i+1}$.

    Next, we show how to construct a function $L_{i+1}$ inductively satisfying the same properties. Indeed, let
    \[
    L_{i+1}(y_1, y_2, \dots y_{i+1}) = L_{i}(y_1, \dots y_{i}) b_{i+1}^{y_{i+1}} L_{i}(y_1, \dots y_{i})^{-1} b_{i+1}^{-y_{i+1}}.
    \]
    Now, let us consider an assignment $(y_1, \dots y_{i+1})$. There are a few cases for us to consider:
    \begin{enumerate}
        \item First, let us suppose that $(y_1, \dots y_{i}) \neq (1, \dots 1)$ (that is the first $i$ bits are not all $1$'s). Then, $L_{i}(y_1, \dots y_{i}) = 0$, and the above expression evaluates to $b_{i+1}^{y_{i+1}} b_{i+1}^{-y_{i+1}} = 1$.
        \item Now, let us suppose that $(y_1, \dots y_{i}) = (1, \dots 1)$, and that $y_{i+1} = 0$. Then, the above expression evaluates to $L_{i}(y_1, \dots y_{i}) L_{i}(y_1, \dots y_{i})^{-1} = 1$. 
        \item Finally, let us suppose that $(y_1, \dots y_{i+1}) = (1, \dots 1)$. Then, the above expression evaluates to
        \[
        L_{i}(1, \dots 1) b_{i+1} L_{i}(1, \dots 1)^{-1} b_{i+1}^{-1}.
        \]
        Recall that by construction, $b_{i+1}$ is an element such that 
        \[
        L_{i}(1, \dots 1) b_{i+1} \neq b_{i+1} L_{i}(1, \dots 1),
        \]
        and as such the above expression is not the identity element. 
    \end{enumerate}

    Thus, our construction satisfies the inductive hypothesis.

    Now, for the function $L_r(y_1, \dots y_r)$, it remains only to show that we can write it in the form described above. First, we show that indeed, the number of monomials that appear in $L_i(y_1, \dots y_r)$ is bounded by $3^{i}$. For this, observe that in our base case $L_2$ only has $4$ monomials. Now, by induction, the number of monomials in $L_{i+1}$ is $2$ times the number of monomials in $L_i$ plus the additional $2$ monomials interspersed. Thus, $L_{i+1}$ has $\leq 2 \cdot 3^i + 2 \leq 3 \cdot 3^i \leq 3^{i+1}$ distinct monomials. 

    Now, the linear equation and map $\pi$ are essentially trivial given the above. We define 
    \[
        C_2(x_1, x_2, x_3, x_4) = b_1^{x_1} b_2^{x_2} b_1^{-x_3} b_2^{-x_4}.
    \] 
    Now, let $K_i$ denote the number of monomials in the constraint of arity $i$. We define the $i$th linear equation as 
    \[
    C_i(x_1, \dots x_{K_{i}}) = C_{i-1}(x_1, \dots x_{K_{i-1}}) b_i^{x_{K_{i-1} + 1}} C_{i-1}(x_{K_{i-1} + 2}, \dots x_{2K_{i-1} + 1})^{-1} b_i^{-x_{2K_{i-1} + 2}}.
    \] Likewise, the first permutation map $\pi_2: \{x_1, \dots x_4\} \rightarrow \{y_1, y_2\}$ simply maps $x_1 \rightarrow y_1, x_2 \rightarrow y_2, x_3 \rightarrow y_1, x_4 \rightarrow y_2$. We can then observe that 
    \[
    C_2(\pi_2(x_1), \pi_2(x_2), \pi_2(x_3), \pi_2(x_4)) = L_i(y_1, y_2).
    \]
    Inductively, we let $\pi_i$ be defined such that for $j \in [K_{i-1}]$, $\pi_i(x_j) = \pi_{i-1}(x_j)$, $\pi_i(x_{K_{i-1}+1}) = y_{i}$, for $j \in [K_{i-1}+2, 2K_{i-1}+1]$, $\pi_i(x_j) = \pi_{i-1}(x_{j - (K_{i-1}+1)})$, and lastly $\pi_i(x_{2K_{i-1}+2}) = y_i$. With this correspondence, we achieve exactly that 
    \[
    C_i(\pi_i(x_1), \dots \pi_i(x_{K_i})) = L_i(y_1, \dots y_i).
    \]
    By letting $i = r$, we then obtain the stated claim.  
\end{proof}

Now, we can conclude the following which proves \Cref{thm:lower-bound-intro}:

\begin{theorem} \label{thm:lower-bound}
    Given the set of variables $x_1, \dots x_n$, and a parameter $\eps > 0$, for any $r = o(\sqrt{\log n})$, there is a set of linear equations $C_1(x_1,\dots x_n), \dots C_m(x_1, \dots x_n)$ with coefficients in $\mathbf{D}_3$ such that any $(1 \pm \eps)$ sparsifier requires preserving $n^{r - o(1)}$ of the equations $C_1, \dots C_m$.
\end{theorem}

\begin{proof}
    This follows by first invoking \cref{clm:simulateAND}, and then invoking Theorem 3.4 of \cite{khanna2025efficient} (stated as \cref{thm:andLB} above). Note that for simulating an AND of arity $r$, we require linear equations over $\leq 3^r$ variables. In turn, this implies that our sparsification lower bound is 
    \[
    \Omega\left ( \left (\frac{n}{3^r} \right )^r  \right ) = \Omega\left ( n^{r - \frac{r^2 \log(3)}{\log n}} \right ).
    \]
    For any choice of $r = o(\sqrt{\log n})$, this gives us a sparsification lower bound of $n^{r - o(1)}$.
\end{proof}